\documentclass[letterpaper]{article}
\usepackage{flairs}

\usepackage{times}
\usepackage{helvet}
\usepackage{courier}
\usepackage{natbib}
\usepackage{hyperref}
\usepackage{graphicx}
\usepackage[table]{xcolor}
\usepackage{enumitem}
\usepackage{multicol}
\usepackage{multirow}
\usepackage{soul}
\usepackage{amssymb}   
\usepackage{amsmath}   
\usepackage{array}
\usepackage{pifont}
\usepackage{makecell}
\usepackage{xspace}
\usepackage{multirow}
\usepackage{tabularx}   
\usepackage[table]{xcolor} 
\usepackage{xcolor}
\usepackage{url}
\usepackage{xurl} 
\usepackage{fancyhdr}
\usepackage{cleveref}

\definecolor{cA}{HTML}{FFB3BA} 
\definecolor{cB}{HTML}{FFDFBA} 
\definecolor{cC}{HTML}{FFFFBA} 
\definecolor{cD}{HTML}{E4F0C3} 
\definecolor{cE}{HTML}{C8E1CC} 

\newcommand{\highlight}[2]{%
  \begingroup
  \sethlcolor{#1}%
  \hl{#2}%
  \endgroup
}
\newcommand{\cA}[1]{\cellcolor{cA}{#1}}
\newcommand{\cB}[1]{\cellcolor{cB}{#1}}
\newcommand{\cC}[1]{\cellcolor{cC}{#1}}
\newcommand{\cD}[1]{\cellcolor{cD}{#1}}
\newcommand{\cE}[1]{\cellcolor{cE}{#1}}

\newcolumntype{L}{>{\raggedright\arraybackslash}p{0.32\columnwidth}}
\newcolumntype{C}{>{\centering\arraybackslash}X}
\newcolumntype{Y}{>{\centering\arraybackslash}X}

\renewcommand{\arraystretch}{1.2}     
\newcolumntype{C}{>{\centering\arraybackslash}X}

\usepackage{tabularx}  
\usepackage{booktabs}  
\usepackage{array}  

\newcommand{\framework}{\textsc{ACE-TA}\xspace}

\begin{document}


\title{\framework: An Agentic Teaching Assistant for Grounded Q\&A, Quiz Generation, and Code Tutoring}



\author{
Himanshu Tripathi$^{\dagger}$,
Charlottee Crowell$^{\dagger}$,
Kaley Newlin$^{\ddagger}$,
Subash Neupane$^{\S}$,
Shahram Rahimi$^{\dagger}$,
Jason Keith$^{\P}$\\[4pt]
$^{\dagger}$University of Alabama, Tuscaloosa, Alabama, USA\\
$^{\ddagger}$Brown University, Providence, Rhode Island, USA\\
$^{\S}$Meharry Medical College, Nashville, Tennessee, USA\\
$^{\P}$Iowa State University of Science and Technology, Ames, Iowa, USA\\[2pt]
\footnotesize
(htripathi, ccrowell)@crimson.ua.edu;
kaley\_newlin@brown.edu;
subash.neupane@mmc.edu;
srahimi@ua.edu;
jkeith@iastate.edu
}


\maketitle
\begin{abstract}
\begin{quote}

We introduce \framework, the Agentic Coding and Explanations Teaching Assistant framework, that autonomously routes conceptual queries drawn from programming course material to grounded Q\&A, stepwise coding guidance, and automated quiz generation using pre-trained Large Language Models (LLMs). \framework consists of three coordinated modules: a retrieval-grounded conceptual Q\&A system that provides precise, context-aligned explanations; a quiz generator that constructs adaptive, multi-topic assessments targeting higher-order understanding; and an interactive code tutor that guides students through step-by-step reasoning with sandboxed execution and iterative feedback.


\end{quote}
\end{abstract}

\section{Introduction} \label{sec:Intro}

Higher education is experiencing sustained growth following the sharp enrollment declines caused by the COVID-19 pandemic. Between Spring 2024 and Spring 2025, post-secondary enrollment in the United States increased by more than half a million students, with undergraduate enrollment rising by 3.5\% \citep{nscrc2020enrollment}. While encouraging, this surge intensifies longstanding instructional challenges: instructors frequently manage multiple large course sections, and Teaching Assistants (TAs) (often students themselves) face substantial workloads and limited availability. As a result, many learners struggle to obtain timely help outside scheduled office hours, particularly when questions arise during homework, late-night study sessions, or while wrestling with complex programming problems. These access constraints can hinder conceptual understanding, reduce motivation, and exacerbate performance gaps in high-enrollment Science, Technology, Engineering, and Mathematics (STEM)  courses.



At the same time, recent advances in Agentic Artificial Intelligence (AAI) offer new opportunities to provide targeted on-demand educational support. Unlike traditional Large Language Model (LLM) systems, agentic frameworks couple language models with capabilities such as Retrieval-Augmented Generation (RAG), tool use, memory, and multi-step planning, enabling them to autonomously decompose tasks, invoke external tools, and operate with minimal human supervision \citep{acharya2025agentic, sapkota2026agentic}.
Despite the capabilities of AAI frameworks, their integration into higher education remains limited. Most existing educational LLM systems focus narrowly on natural-language Q\&A or administrative support \citep{khurana2023nlp}, whereas agentic systems can leverage planning, tool use, and retrieval to support computational tasks that require multi-step reasoning, including code construction, debugging, and stepwise problem-solving. These are precisely the areas where novice programmers typically need the most guidance. For example, students in introductory programming courses often grasp concepts abstractly but struggle to implement them in code, debug errors, or reason about program behavior. Existing systems improve access to course information, but offer limited assistance with applied problem-solving or self-evaluation \citep{keith2024barkplug1, neupane2024barkplug, keith2025barkplug}. This creates a need for unified systems that support conceptual understanding, guided code construction, and formative assessment.

To address this need, we introduce the Agentic Coding and Explanations Teaching Assistant (\framework), an AAI framework that supports students in conceptual reasoning, guided code development, and formative assessment through adaptive quizzes. \framework incorporates three synergistic components: (1) a retrieval-grounded conceptual Q\&A module that delivers concise, textbook-aligned explanations; (2) an adaptive quiz generator that constructs higher-order, multi-concept assessments and tailors difficulty based on learner responses; and (3) an interactive stepwise code tutor that decomposes programming problems, validates student code through sandboxed execution, and provides iterative feedback. Together, these capabilities allow \framework to serve as an after-hours teaching assistant that students can consult for conceptual clarification, structured practice, and guided problem-solving.

The main contributions of this paper are as follws: \begin{itemize}
\item 
We design an agentic teaching assistant (\framework) that provides grounded conceptual Q\&A, adaptive quiz generation, and real-time stepwise coding guidance using pre-trained LLMs.

\item We introduce novel integrations of hybrid retrieval, multi-concept quiz generation, and incremental code tutoring with sandboxed execution and iterative feedback.

\item We demonstrate the effectiveness of \framework 
using a combination of quantitative metrics 
and expert human evaluation. 


\end{itemize}


The remaining sections of this work are structured as follows. Section \hyperlink{sec:Architecture}{2} describes the ACE-TA architecture in greater detail, outlining the orchestration layer and specific modules for Q\&A, quiz generation, and code tutoring. Section \hyperlink{sec:Results}{3} provides an analysis of its performance using both quantitative metrics and qualitative Subject Matter Expert (SME) evaluations. Section \hyperlink{sec:RelatedWorks}{4} benchmarks the framework against representative architectures and discusses related works. Finally, in Section \hyperlink{sec:ConclusionFutureResearch}{5}, we conclude the work and discuss future research opportunities.

\hypertarget{sec:LitRev}{}

\hypertarget{sec:Architecture}{}
\section{Architecture \& Method}

\begin{figure*}[!htbp]
\vspace{-4pt}
\centering

\centering
\includegraphics[scale=0.60]{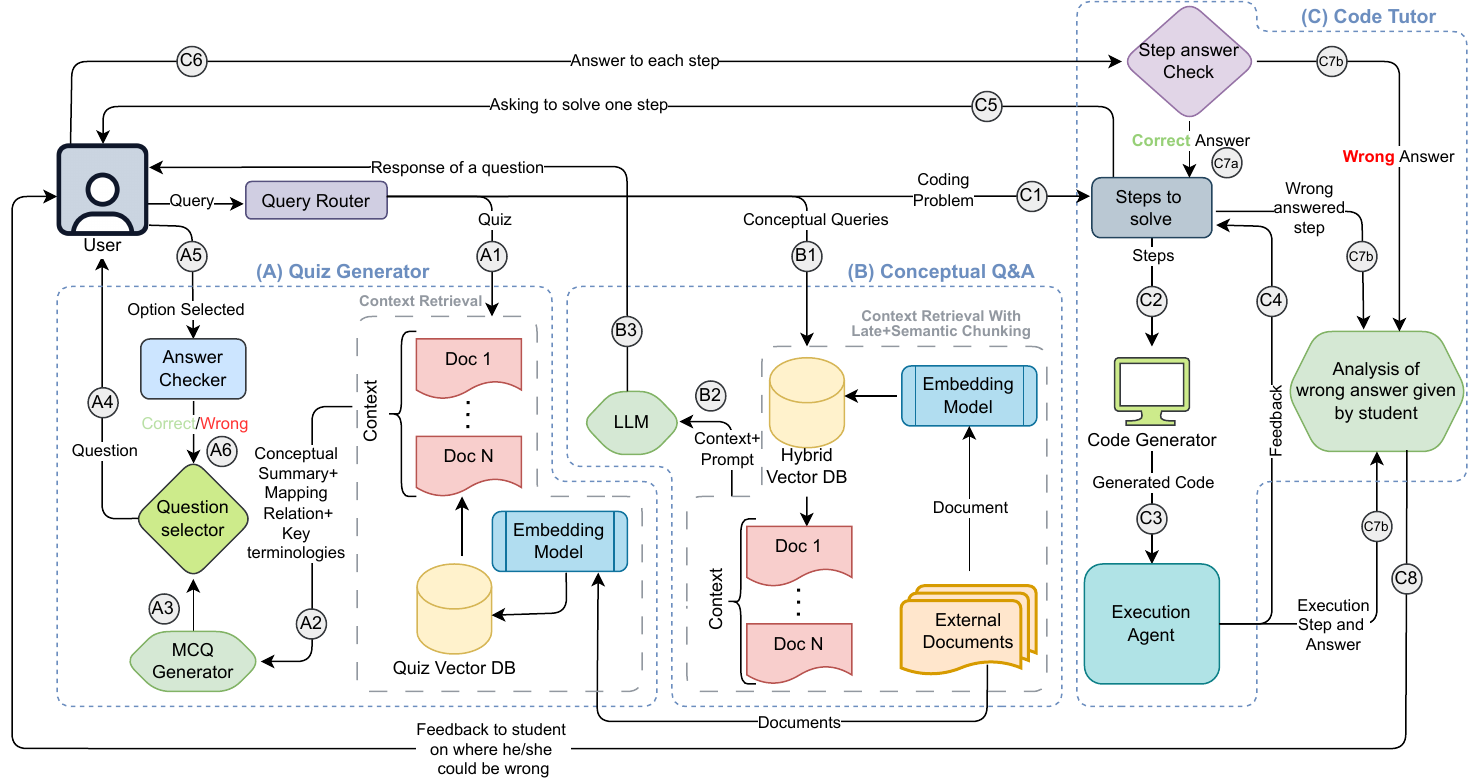}

\caption{\framework multi agent workflow for routed retrieval, quiz construction, and stepwise, feedback driven code tutoring.}
\vspace{-6pt}
\label{AssignmentCoach}
\end{figure*}           

\framework is a local, multi-module tutoring framework that integrates RAG, adaptive assessment, and stepwise code coaching under a shared orchestration layer as depicted in Fig.  \ref{AssignmentCoach}. The framework routes each learner query to one of three specialized pathways such as \textit{Quiz Generator, Conceptual Q\&A,} or \textit{Code Tutor}, while reusing shared LLM and retrieval resources to minimize initialization overhead.

\subsection{Orchestration Layer (Query Router)}
A lightweight orchestration layer implements query routing and module dispatch. Two local LLMs are loaded via $llama-cpp-python$: a compact router/validator model $Phi-3-Mini-128K-Instruct\ (Q4\_K\_M)$ \footnote{\url{https://huggingface.co/microsoft/Phi-3-mini-128k-instruct}} and a larger generator model $GPT-OSS-20B\ (Q4\_K\_S)$ \citep{gptoss20b}. The router is prompted with a constrained schema listing available tools and representative query patterns, and outputs exactly one label from {Conceptual Q\&A, Quiz Generator, Code Tutor, Unknown}. Routing is executed deterministically (temperature $0$). The orchestrator then dispatches the original query to the selected module while reusing preloaded model objects to reduce per-query latency.


\subsection{Conceptual Query handling}
Conceptual question (Fig. \ref{AssignmentCoach}, \texttt{(B1)}) is answered via hybrid retrieval, reranking, and constrained generation over the Hybrid Vector DB. \framework executes FAISS-based dense retrieval and BM25 lexical retrieval, then forms a candidate pool as the union of the top-K results from both channels ($K=20$). Then, candidate passages are reranked using a cross-encoder $MS\ MARCO-MiniLM-L-6-v2$ \footnote{\url{https://huggingface.co/cross-encoder/ms-marco-MiniLM-L6-v2}}, and the top five passages are selected as the final context. Finally, the selected passages are concatenated into a context block and injected into a structured Harmony style prompt \footnote{\url{https://cookbook.openai.com/articles/openai-harmony}} (system: tutoring role; developer: constraints + retrieved context; user: query). The generator is instructed to condition responses (Fig. \ref{AssignmentCoach}\texttt{(B3)}) on retrieved evidence (Fig. \ref{AssignmentCoach}\texttt{(B2)}) and to explicitly acknowledge insufficient support when the context is inadequate.

\subsection{Quiz Generator}

The quiz module generates Multiple Choice Quiz (MCQ) targeting higher-order understanding and supports adaptive difficulty. For broad requests (e.g., asking for a quiz on a Python function), the validator ($Phi-3-Mini-128K-Instruct\ (Q4\_K\_M)$) decomposes the topic into five candidate subtopics (JSON list) for learner selection, reducing topic drift and improving item specificity (Fig. \ref{AssignmentCoach} \texttt{(A1)}). The refined topic embedding (all-MiniLM-L6-v2\footnote{\url{https://huggingface.co/sentence-transformers/all-MiniLM-L6-v2}}) retrieves $50$ candidate chunks from the Quiz Vector DB. Then we utilize Maximum Marginal Relevance (MMR) that selects $25$ diverse passages:

\begin{equation*}
    MMR(c_i) = \lambda sim(q, c_i) - (1 - \lambda) \max_{c_j \in S} sim(c_i, c_j)
\end{equation*}

\noindent where, $q$ is the query, $c_i, c_j$ are candidate chunks, $S$ is the set of already-selected diverse chunks and $\lambda$ is a trade-off parameter ($\lambda = 0.7$) (Fig. \ref{AssignmentCoach}, \texttt{(A2)}). The generator constructs a compact concept framework (e.g., mechanisms, misconceptions, constraints, trade-offs) (Fig. \ref{AssignmentCoach}, \texttt{(A3)}) and synthesizes scenario-based MCQs with four options, a labeled correct answer, distractor rationales, Bloom-level tags (Apply/Analyse/Evaluate/Create), and the concepts assessed. Then, the validator ($Phi-3-Mini-128K-Instruct\ (Q4\_K\_M)$) checks structural validity, Bloom alignment, and multi-concept coverage, filtering low-quality items prior to presentation (Fig. \ref{AssignmentCoach}, \texttt{(A4)}). For adaptive progression (Fig. \ref{AssignmentCoach}, \texttt{(A5, A6)}), the difficulty of the subsequent item is adjusted based on correctness, promoting items of higher-order upon correct responses, and providing corrective feedback with lower-level reinforcement otherwise.

\subsection{Code Tutor}

The Code Tutor provides stepwise guidance for translating natural-language questions (Fig. \ref{AssignmentCoach}, \texttt{(C1)}) into executable Python programs. First, it invokes a planning prompt for the generation of a plan utilizing $GPT-OSS-20B\ (Q4\_K\_S)$ as a generator. The generator produces a numbered sequence of logic-level steps (no code, no environment instructions) that decomposes the task into minimal reasoning units. For each step the tutor builds a second prompt that contains the cumulative code written so far and the current step description (this is excluded for the first step where only current step description is provided) and asks the LLM for the smallest new Python snippet that implements only that step (Fig. \ref{AssignmentCoach}, \texttt{(C2)}), which is then executed (Fig. \ref{AssignmentCoach}, \texttt{(C3)}) and if failed it gives feedback back to modify the steps (Fig. \ref{AssignmentCoach}, \texttt{(C4)}). which prevents premature inclusion of later logic and clarifies the boundary between ideas. The learner sees the textual step (Fig. \ref{AssignmentCoach}, \texttt{(C5)}), reviews the current code (if they have written something), and then types only the new lines they believe are needed (Fig. \ref{AssignmentCoach}, \texttt{(C6)}), while the tutor handles boilerplate such as appending a minimal body when a line ends with a block opener. The student code first passes through an Abstract Syntax Tree called AST to catch syntax or indentation issues before pushing it to the sandbox testing. This ensures syntax and structural errors are caught early, avoiding wasted sandbox runs and giving students faster, safer, and more focused feedback. If the user code passes the sandbox test successfully, the system then compares the learner snippet to the reference snippet with an LLM based comparator that focuses on the current step, accepts alternate but equivalent logic, and returns either a confirmation with the next step (Fig. \ref{AssignmentCoach}, \texttt{(C7a)}) or for wrong answers, analyzes the main flaw (Fig. \ref{AssignmentCoach}, \texttt{(C7b)}) and providing a feedback with a short explanation of what the learner would have thought and how they can improve it (Fig. \ref{AssignmentCoach},\texttt{(C8)}). This process yields a complete executable solution built through tightly guided iterations. The same local implementation groundwork then supports the experimental study in which we evaluate \framework across coding support, conceptual Q\&A, and quiz generation tasks.


\hypertarget{sec:Results}{}
\section{Evaluation and Results}




We evaluate \framework using quantitative metrics and qualitative evaluation by Subject Matter Experts (SMEs) including course instructors, teaching assistants, and graduate students, and benchmark against representative architectures. From \textit{Mark Lutz, Learning Python (Fifth Edition), O'Reilly Media, 2013}\footnote{\url{https://www.oreilly.com/library/view/learning-python-5th/9781449355722/}} we curated three datasets: 100 conceptual Q\&A pairs with reference answers and retrieved contexts, 108 MCQs across 35 topics, and 150 coding problems split evenly across difficulty levels, with qualitative subsets rated by SMEs (for more dataset details see \hyperlink{apx:dataset}{Dataset Description and Preparation in Appendix}). \framework employs dual vector databases optimized for distinct retrieval objectives (more details see \hyperlink{apx:VD}{Vector Database in Appendix}). All experiments were conducted on a workstation with GeForce RTX 5090 32 GB, Intel Core Ultra 9 285K CPU, and 64 GB RAM.

\subsection{Retrieval and Answer Fidelity}
\begin{figure} [ht]
    \centering
    \includegraphics[width=0.95\linewidth]{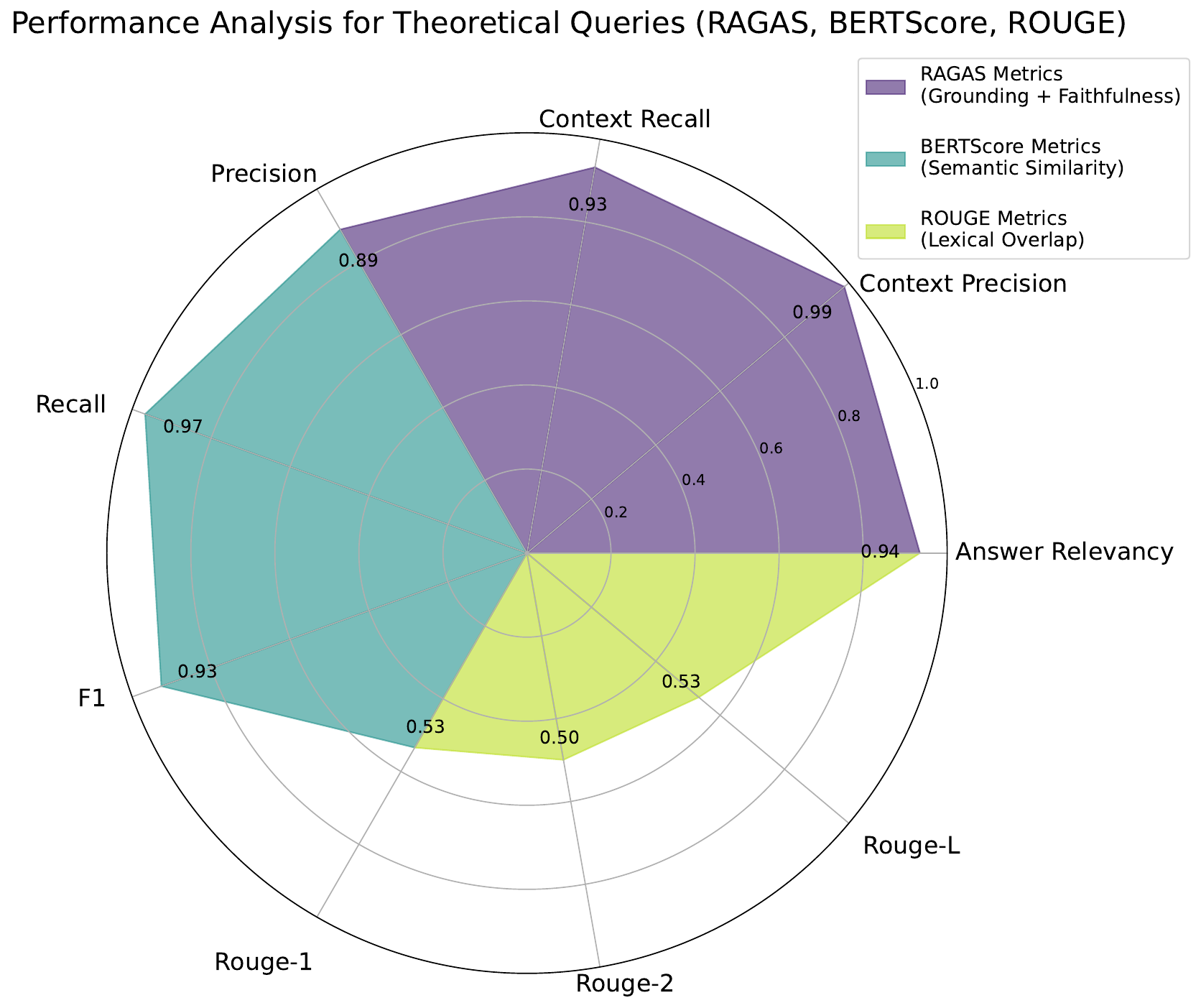}
    \caption{Quantitative performance of the conceptual QA module on natural language queries, showing RAGAS (answer relevancy, context precision/recall), BERTScore (precision/recall/F1), and ROUGE (ROUGE1, ROUGE2, ROUGEL) metrics for retrieval and answer fidelity.}
    \label{fig:theo_quant}
\end{figure}

We first quantify how well the conceptual question answering module performs when it answers Python questions using a hybrid retrieval augmented generation pipeline over textbook material. Using $100$ question answer pairs with relevant chunk retrieved and ground truth generated using Gemini 2.5 Pro, we compute RAGAS answer relevancy and context scores together with BERTScore based on RoBERTa-large and ROUGE to capture both semantic and lexical fidelity. Retrieval shows near perfect context precision with high recall, indicating that the hybrid BM25 and FAISS search plus cross encoder reranker over late chunking and semantic chunking based vector DB reliably surfaces the right passages. Fig. \ref{fig:theo_quant} shows high answer relevancy around $0.94$ and strong BERTScore F1 around $0.93$ suggest very small meaning drift, while moderate ROUGE values (ROUGE1, ROUGE2, ROUGEL $\approx 0.5$) show that the model tends to paraphrase instead of copying, which is helpful for learning. Running at temperature $0$ further supports grounded answers. Overall, these trends indicate that students receive responses that follow textbook intent while using simpler language. Next, we add expert ratings that compare these explanations with those from Gemini 2.5 Pro.

\subsection{SME Judgements and Ground Truth Comparison}

\begin{figure}[ht]
    \centering
    \includegraphics[width=1\linewidth]{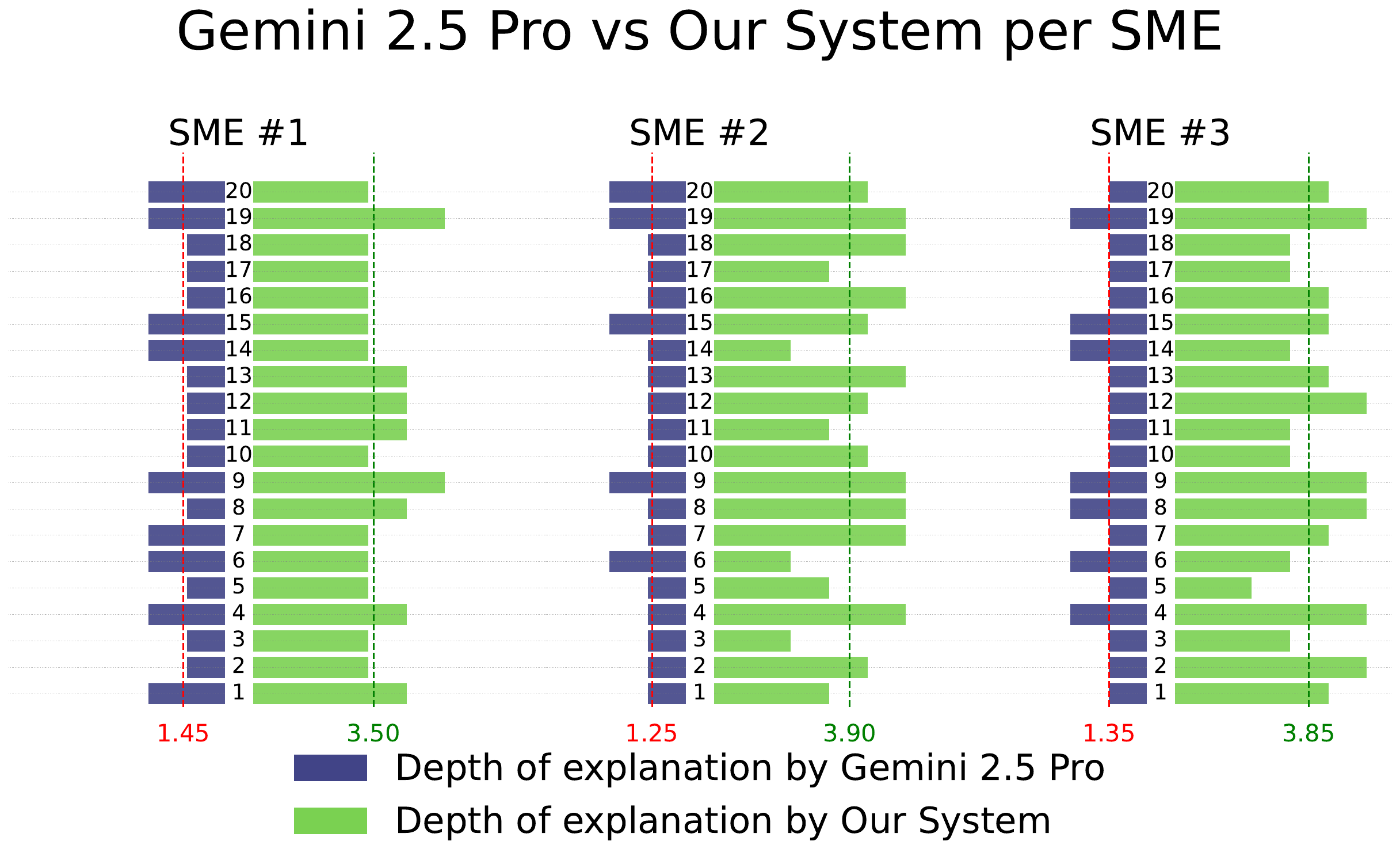}
    \caption{Depth of explanation ratings from three SMEs for Gemini 2.5 Pro and \framework conceptual QA module across 20 questions, showing consistently higher depth scores for \framework where higher values reflect more deeper explanations.}
    \label{fig:theo_quali}
\end{figure}

To examine explanation quality beyond quantitative metrics, we asked 3 SMEs to compare masked answers from \framework and Gemini 2.5 Pro on 20 conceptual Python questions, rating depth of explanation from 1 to 5. Across all 3 experts \framework scores around 3.8 on depth while Gemini 2.5 Pro remains near 1.3 as shown in Fig. \ref{fig:theo_quali}. This pattern suggests that our explanations are not only preferred on average but also reliably judged deeper and more focused. SMEs note that our answers stay on topic, whereas Gemini often introduces side information that feels like clutter, and one SME flags a confusing keyword argument name that could mislead learners. 

\subsection{Topic Breadth and Balance of Quiz Generator}
\hypertarget{apx:subdist}{}
\begin{figure}
    \centering
    \includegraphics[width=1\linewidth]{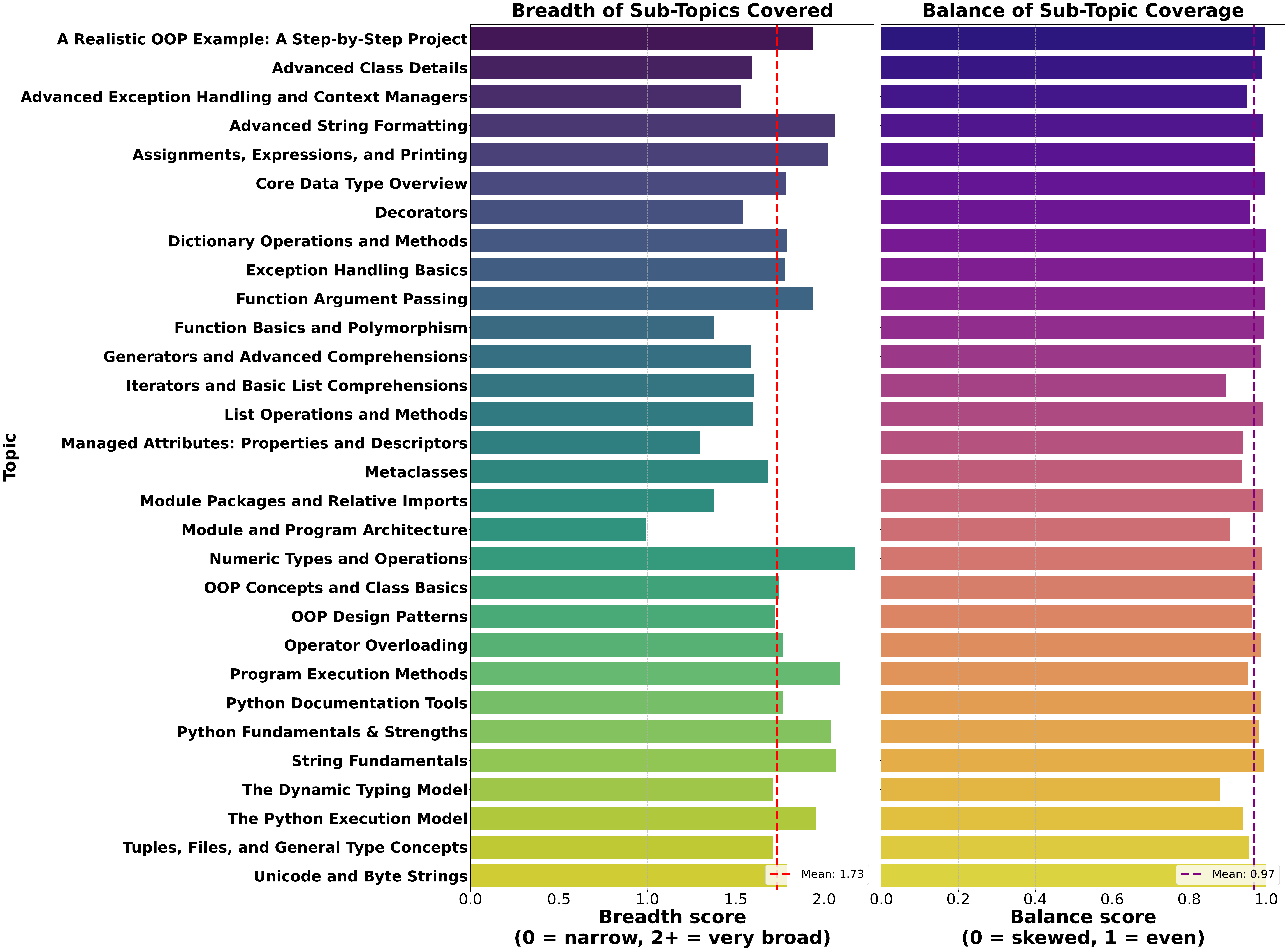}
    \caption{(A) Breadth of quiz subtopic coverage across Python topics, measured via penalised Shannon’s entropy. (B) Balance of subtopic coverage, measured via Pielou’s evenness.}
    \label{fig:quiz_quant}
\end{figure}

We evaluate the \framework quiz generator using distribution based metrics that quantify how broadly and uniformly each topic's conceptual space is sampled. The topics and subtopics are derived from Mark Lutz’s Learning Python, and GPT OSS 20B assigns subtopic relevance scores for 108 generated questions spanning 35 topics. From these scores, we compute two summary measures per topic that capture coverage and balance.


\begin{description}
    \item[Penalised Shannon's entropy] for breadth is given by
    \begin{equation*}
        H_{pen} = - \sum_1^k p_i \ln p_i
    \end{equation*}
    where $H_{pen}$ is the penalised Shannon's entropy for a given topic (reduced when its subtopics are heavily shared with other topics, so redundant coverage counts less), $p_i$ is the normalised relevance weight of subtopic $i$, and $k$ is the total number of subtopics for that topic (Fig. \ref{fig:quiz_quant} (A)).
    \item[Pielou's evenness] is defined as:
    \begin{equation*}
        J = \frac{H_{pen}}{\log (subtopic\ count)}
    \end{equation*}
    where, $J$ is Pielou’s evenness index for a topic, which measures how evenly the total relevance is spread across its $subtopic\ count$ (Fig. \ref{fig:quiz_quant} (B)).
\end{description}


Breadth scores span $0.5$ to $2.17$ (mean $1.73$), with topics like ``Numeric Types and Operations" showing wide coverage. Evenness remains high ($0.88$ to $1.00$, mean $0.97$), indicating no single subtopic dominates. Fig. \ref{fig:quiz_quant} shows topics achieve balanced coverage without extreme skew. No topic falls into the extreme regime of simultaneously low entropy and high skew. This indicates that the retrieval stage, conceptual framework extraction, and MCQ synthesis collectively promote quizzes that are broad in coverage and reasonably balanced between subtopics.


\subsection{Explanation Adequacy of the Quiz Generator from SME Responses}

\begin{figure}[ht]
    \centering
    \includegraphics[width=0.85\linewidth]{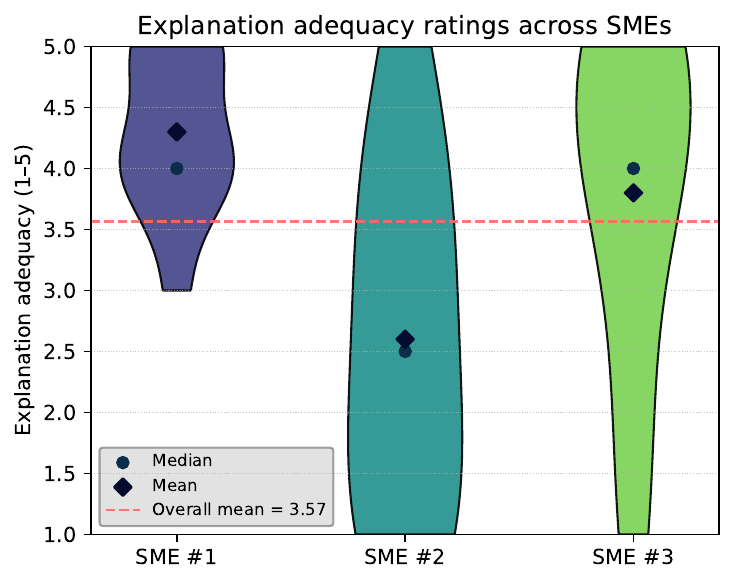}
    \caption{Explanation adequacy ratings from three SMEs for quiz master explanations across 10 questions, showing SME specific means and medians with an overall mean of 3.57, indicating generally adequate but improvable explanatory quality. }
    \label{fig:quiz_quali}
\end{figure}

Three SMEs rated explanations for 10 generated questions on a five point scale (1 poor, 5 very strong). Fig. \ref{fig:quiz_quali} shows mid to upper ratings with an overall mean of 3.57, suggesting explanations are generally adequate but variable. SMEs noted occasional misalignment between question wording and rationale, and reliance on unstated assumptions in options, which we partly attribute to token limit constraints when GPT OSS 20B generates multiple items and explanations in a single pass. We next evaluate Code Tutor support during programming practice.


\subsection{Robustness and Solution Quality of Coding Tutor}
\begin{table}[t]
\centering
\caption{
Evaluation of code-generation models across Output Match Rate, CodeBERT similarity, and CodeBLEU for easy, medium, and hard Python coding tasks. 
\textit{Values are color–coded into five ranges:
\highlight{cA}{0.100--0.299},
\highlight{cB}{0.300--0.499},
\highlight{cC}{0.500--0.699},
\highlight{cD}{0.700--0.899},
\highlight{cE}{0.900--1.000}.
}}

\label{tab:code_metrics}
\renewcommand{\arraystretch}{1.25}

\resizebox{\columnwidth}{!}{%
\begin{tabular}{p{3.5cm}|ccc|ccc|ccc}
\hline
\rowcolor{white!20!}
\textbf{Model} &
\multicolumn{3}{c|}{\textbf{Output Match Rate}} &
\multicolumn{3}{c|}{\textbf{CodeBert (sim\_max)}} &
\multicolumn{3}{c}{\textbf{CodeBLEU (codebleu)}} \\
\rowcolor{white!20!}
 & Easy & Medium & Hard
 & Easy & Medium & Hard
 & Easy & Medium & Hard \\
\hline

\cellcolor{white!20!}Gemma-3-27B-IT (Q4\_0)
 & \cE{1.000} & \cE{1.000} & \cE{1.000}
 & \cE{0.993} & \cE{0.994} & \cE{0.990}
 & \cC{0.500} & \cC{0.544} & \cB{0.411} \\ \hline

\cellcolor{white!20!}Llama-4-Scout-17B-16E-Instruct
 & \cE{1.000} & \cE{1.000} & \cE{1.000}
 & \cE{0.992} & \cE{0.995} & \cE{0.992}
 & \cB{0.489} & \cC{0.510} & \cB{0.429} \\ \hline

\cellcolor{white!20!}GPT-4o-Mini
 & \cE{0.980} & \cE{1.000} & \cE{0.960}
 & \cE{0.993} & \cE{0.994} & \cE{0.992}
 & \cB{0.465} & \cC{0.516} & \cB{0.409} \\ \hline

\cellcolor{white!20!}DolphinCoder-StarCoder2-15B (Q8\_0)
 & \cE{1.000} & \cE{0.980} & \cE{0.940}
 & \cE{0.994} & \cE{0.994} & \cE{0.989}
 & \cC{0.512} & \cB{0.487} & \cB{0.415} \\ \hline

\cellcolor{white!20!}DeepSeek-R1-Qwen3-8B-BF16
 & \cE{1.000} & \cE{0.980} & \cE{0.940}
 & \cE{0.994} & \cE{0.995} & \cE{0.989}
 & \cC{0.557} & \cC{0.589} & \cB{0.404} \\ \hline

\cellcolor{white!20!}Maverick-7B (FP16)
 & \cE{1.000} & \cE{0.960} & \cE{0.960}
 & \cE{0.995} & \cE{0.995} & \cE{0.990}
 & \cC{0.546} & \cC{0.565} & \cB{0.403} \\ \hline

\cellcolor{white!20!}Mistral-7B-Instruct-v0.2 (8-bit)
 & \cE{0.920} & \cD{0.860} & \cC{0.660}
 & \cE{0.983} & \cE{0.979} & \cE{0.974}
 & \cB{0.336} & \cB{0.316} & \cA{0.243} \\ \hline

\end{tabular}
} 
\end{table}

\begin{figure}
    \centering
    \includegraphics[width=1\linewidth]{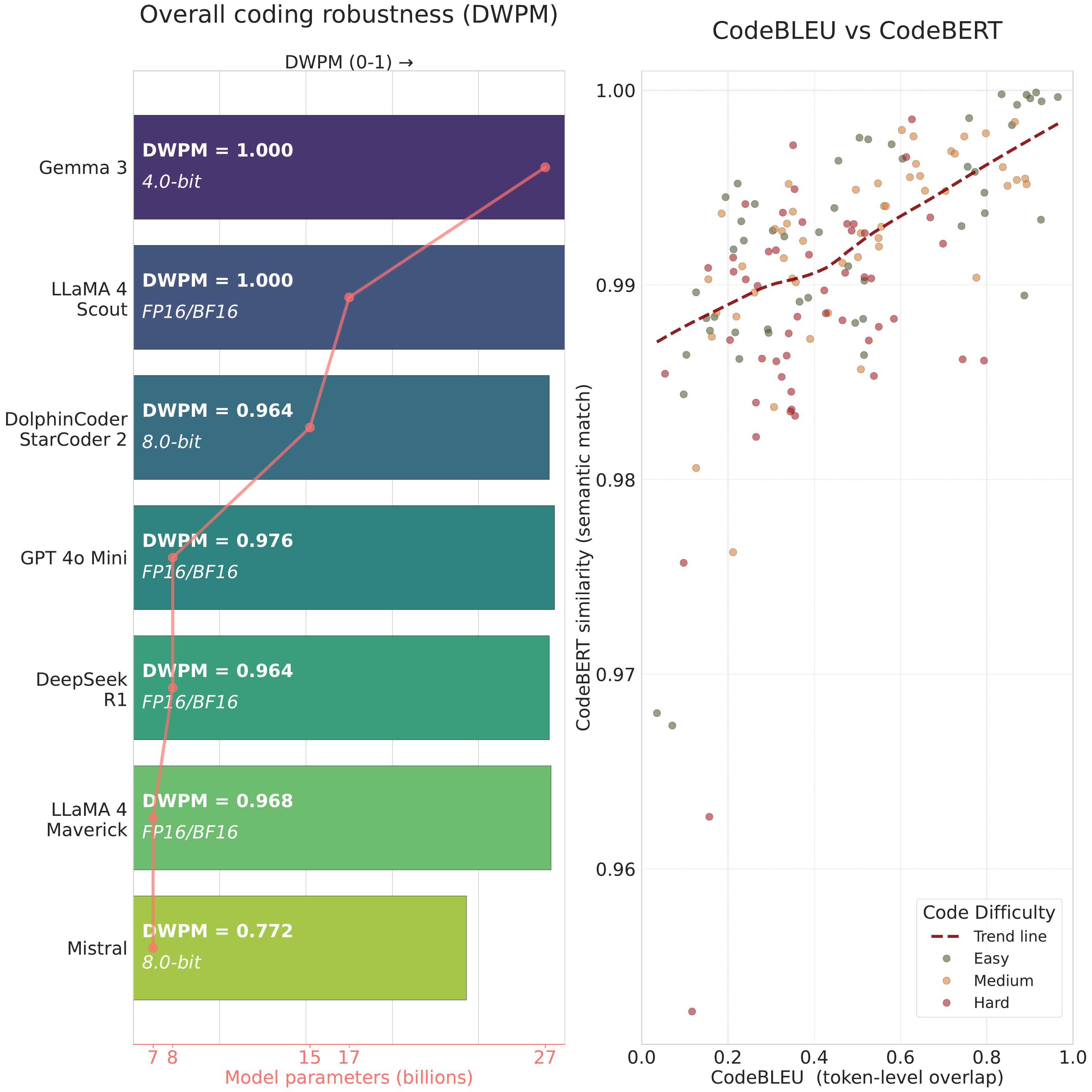}
    \caption{(A) Overall coding robustness with model size and quantization of 7 student LLMs measured by DWPM. (B) CodeBLEU vs. CodeBERT similarity across problems, coloured by difficulty level. }
    \label{fig:code_quant}
\end{figure}

We evaluate the Code Tutor by measuring robustness across difficulty levels and code quality utilizing seven models (see Table \ref{tab:code_metrics}) spanning 7B to 27B parameters and 4 bit to BF16 precision, operated inside the tutor loop. Each step allows up to five refinement attempts after sandbox feedback, and each problem allows up to three final sandbox attempts. We evaluate 150 problems (50 easy, 50 medium, 50 hard). We summarize task robustness using a Difficulty Weighted Performance Measure (DWPM):



\[
\begin{aligned}
\mathrm{DWPM} &= 0.2A_{\text{easy}}+0.3A_{\text{med}}+0.5A_{\text{hard}},\\
A_d &= \frac{C_d}{50},\quad d\in\{\text{easy},\text{med},\text{hard}\}.
\end{aligned}
\]

\noindent where, the weighting (e.g., 0.2, 0.3, 0.5) emphasizes harder problems, reflecting their longer and more compositional step sequences (weights set by an SME) and $C_d$ is the number of problems solved correctly at difficulty $d$.







Using identical task prompts (adapted only to each model’s native template), Fig. \ref{fig:code_quant}(A) shows that larger parameter models (e.g., Gemma 3 27B, Llama 4 Scout 17B) achieve near ceiling DWPM, whereas quantised models such as Mistral 7B 8 bit  degrade primarily due to formatting deviations that trigger sandbox failures (for example, adding non code prefaces and conclusions despite tutor warnings).


To quantify the association between model capacity and DWPM, we combine parameter count and numeric precision into an effective capacity proxy:

\begin{equation*}
    Effective\ Size_i = params_i + \frac{precision_i}{10}
\end{equation*}



\noindent where, $i$ is LLM, $params_i$ is the number of parameters, and $precision_i$ is the numerical precision level for that model. The division by 10 rescales precision so that it modulates but does not dominate the parameter term. Under this proxy, the models have $Effective\ Size_i$ values of 27.40 (Gemma 3 27B Q4), 18.60 (Llama 4 Scout 17B 16E), 16.60 (GPT 4o Mini), 15.80 (DolphinCoder StarCoder2 15B Q8), 9.60 (DeepSeek Qwen3 8B BF16), 8.60 (Maverick 7B FP16), and 7.80 (Mistral 7B 8 bit). We then compute the Spearman rank correlation between model capacity and DWPM. Models are ranked by $Effective\ Size_i$ (ascending) and independently ranked by DWPM (ascending), using average ranks for ties. The resulting Spearman correlation is
\[
\rho \;=\; 1 - \frac{6\sum_{i=1}^{n} d_i^2}{n\left(n^2-1\right)}
\;=\; 0.875 \quad (n=7),
\]
indicating a strong positive association between effective capacity and difficulty weighted robustness for student models operating inside the tutor architecture.

Furthermore, to characterize solution quality beyond pass or fail, we compute CodeBLEU and CodeBERT similarity and report means over all 150 problems per model (Fig. \ref{fig:code_quant}(B), Table \ref{tab:code_metrics}). Across models, CodeBERT scores are consistently high while CodeBLEU is moderate, reflecting semantic correctness with lexical variation relative to the reference. The small deviation in the trend line is explained by a subset of problems (9 total - 3 easy, 4 medium, and 2 hard) where models produce concise, semantically equivalent solutions using built in functions or alternative control structures, which the tutor appropriately accepts. These results suggest that the tutor’s stepwise decomposition, iterative correction, and sandbox execution enable diverse LLMs to converge to runnable and logically sound code. While higher capacity yields higher DWPM, smaller models still improve through repeated iterations, implying that robustness is driven both by model strength and by the tutor’s structured guidance and verification.

\subsection{SME Judgements on Depth of Coding Tutor}

\begin{figure}[ht]
    \centering
    \includegraphics[width=1\linewidth]{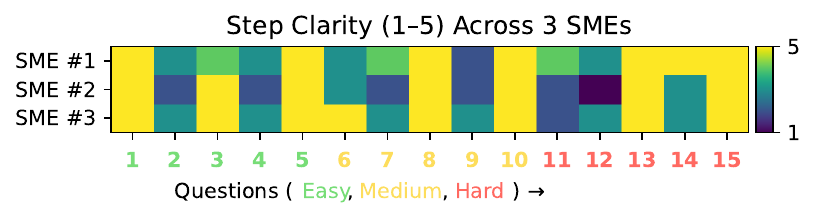}
    \caption{Step clarity ratings from three SMEs across 15 coding tutor questions grouped by easy medium and hard}
    \label{fig:coding_quali}
\end{figure}

To assess step clarity in the coding tutor, we asked three SMEs to rate tutor-generated step sequences for 15 coding questions on a five-point Likert scale, where 1 denotes very unclear and 5 denotes very clear (Figure~\ref{fig:coding_quali}). 
For each SME, we compute the mean rating as $\bar{x}_{k}=\frac{1}{15}\sum_{j=1}^{15}x_{k,j}$, where $x_{k,j}$ is the rating assigned by SME $k\in\{1,2,3\}$ to question $j$. The resulting means are 4.07, 3.47, and 4.00, indicating that SMEs generally find the step sequences clear and usable. We then compute the overall mean across SMEs as $\bar{x}_{\mathrm{all}}=\frac{1}{3}\sum_{k=1}^{3}\bar{x}_{k}\approx 3.84$, which suggests that the tutor typically produces well-structured, instructionally useful steps. Finally, we stratify ratings by difficulty to examine robustness under increasing task complexity. Mean step clarity is 4.00 for easy, 3.80 for medium, and 3.73 for hard problems, showing only a modest decline as difficulty increases. SME comments are consistent with these trends: steps are usually readable, but occasionally rely on basic programming terminology that may be unfamiliar to absolute beginners. In practice, this implies learners receive a clear, ordered solution path, while instructors can further improve accessibility by adding brief glossaries or short reminders for key terms.

\hypertarget{sec:RelatedWorks}{}
\section{Related Works and Comparison Study}

LLM-based teaching assistants for educational Q\&A and tutoring have been explored extensively. BarkPlug \cite{keith2024barkplug1, neupane2024barkplug} leveraged RAG to provide grounded answers about campus resources, extended to BarkPlug 2.0 \cite{keith2025barkplug} as a multi-agent system for high DFW courses. \citet{hicke2023ai} explored SFT, RAG, and DPO for AI TA course Q\&A. Khanmigo \cite{khanmigo} employs Socratic tutoring, while Pensieve Discuss \cite{yang2025pensievegrader} targets collaborative programming through synchronized editing. Jill Watson \cite{taneja2024jillwatson} and Parrot \cite{dai2025parrot} focus on direct response generation with retrieval grounding.

Despite these advances, several gaps remain. First, most systems address either conceptual Q\&A or coding support in isolation, lacking unified frameworks that integrate both with formative assessment; \framework addresses this through a single agentic pipeline with intelligent routing across three specialized modules. Second, existing retrieval approaches employ generic vector stores rather than task-optimized indexing; \framework introduces dual vector databases with task-specific chunking strategies (late chunking for Q\&A fidelity, semantic-only for quiz diversity). Third, interactive coding support is limited, with systems either providing complete solutions or minimal scaffolding; \framework implements stepwise decomposition with sandboxed execution and iterative feedback at the individual step level. Finally, few systems generate adaptive quizzes for higher-order assessment; \framework produces Bloom-tagged MCQs with difficulty adaptation based on learner responses. For detailed comparison across pedagogical and technical dimensions see \hyperlink{apx:comparison}{Comparison with Existing Work in Appendix}.

\hypertarget{sec:ConclusionFutureResearch}{}
\section{Conclusion \& Future Research Opportunities} 
In this work, we introduced \framework, an agentic teaching assistant that unifies three workflows, grounded conceptual Q\&A, conceptual quiz generation, and stepwise code tutoring. A lightweight router dispatches learner queries to specialized modules for retrieval augmented generation and explanation, diagnostic multiple choice assessment, and sandboxed stepwise code construction, reducing context switching while keeping outputs anchored in course materials. Empirically, \framework retrieves relevant evidence and produces faithful explanations, achieves broad and balanced quiz coverage, and supports robust code completion across diverse models, with SMEs rating both conceptual depth and step clarity favorably. Overall, these results indicate that \framework can provide reliable assistance that strengthens conceptual understanding and practical coding skill. 
Future work will focus on two directions. First, we will evaluate \framework on additional courses beyond a single Python textbook to test generalization. Second, we will conduct a classroom deployment study to measure learning impact and identify failure modes that guide targeted improvements in grounding and feedback quality.

\section*{Acknowledgment}
This work was supported by the PATENT Lab (Predictive Analytics and Technology Integration Laboratory) at the Department of Computer Science, University of Alabama, Tuscaloosa.

     \bibliographystyle{flairs} 
        \bibliography{ref} 

\section{Appendix} 
\label{Appendix}

\subsection{Dataset Description and Preparation}
\hypertarget{apx:dataset}{}

For our experiments, we utilize \textit{Mark Lutz, Learning Python (Fifth Edition), O'Reilly Media, 2013}\footnote{\url{https://www.oreilly.com/library/view/learning-python-5th/9781449355722/}} as our primary data source to evaluate \framework. We curate three datasets. First, the \emph{Conceptual Q\&A} dataset includes 100 textbook questions with their reference answers generated by Gemini 2.5 Pro \citep{gemini25pro}, retrieved context chunks from the book, and corresponding \framework answers, so each entry couples a question, evidence, and two candidate explanations. A qualitative subset of 20 of these questions keeps the same context but stores two masked answers from \framework and Gemini 2.5 Pro together with depth ratings from three SMEs and short comments on focus and clarity. Second, the \emph{Quiz Generator} dataset is designed by prompting the quiz module to produce 108 MCQs across 35 Python topics. For each item, we record the associated subtopic label and its relevance score produced by GPT OSS 20B, with subtopics drawn from the textbook’s topic structure. A qualitative subset of 10 items is rated by SMEs on a five point adequacy scale, with short notes on whether explanations are clear, confusing, or assume unstated prior knowledge. Finally, the \emph{Code Tutor} dataset covers programming support using 150 coding problems, evenly split into easy, medium, and hard. We evaluate seven student facing LLM configurations within the tutor loop; for each model problem pair, we log task success and the number of additional attempts required. For qualitative analysis, SMEs review 15 problems and assign step clarity ratings, noting instructions that are vague or overly technical.

\subsection{Vector Database}
\hypertarget{apx:VD}{}
\framework maintains two vector indices, \textit{Hybrid Vector DB} and \textit{Quiz Vector DB}, each optimized for a distinct retrieval objective instead of a single generic store. For conceptual Q\&A, Hybrid Vector DB is built from course textbook PDFs using late \citep{gunther2024late} and semantic chunking to preserve sentence level coherence while controlling chunk length. Each chunk is indexed with a dense channel using Gemma 2 Embedding 300M \citep{gemma2embedding300m} and FAISS \citep{douze2024faiss}, and a lexical channel using BM25 \citep{robertson1976bm25} over the same text. This hybrid design supports grounded explanations where lexical fidelity and semantic proximity are jointly required. For quiz generation, Quiz Vector DB applies semantic chunking without late alignment to increase concept diversity per chunk, prioritizing coverage and compositional variety over paragraph level reconstruction.


\subsection{Comparison with Existing Work}
\hypertarget{apx:comparison}{}  
\begin{table}[ht]
\centering
\scriptsize
\setlength{\tabcolsep}{0.01pt} 
\caption{Feature comparison across LLM-based tutoring systems. 
$\checkmark$ = full support, $\bullet$ = partial support, $\times$ = not supported.}
\label{tab:ablation}
\renewcommand{\arraystretch}{1}
\begin{tabularx}{\columnwidth}{p{1.7cm}CCC>{\centering\arraybackslash}p{0.75cm}C}
\hline
\rowcolor{cyan!20!}
\textbf{Feature} &
\textbf{\hyperlink{cite.khanmigo}{Khanmigo}} &
\textbf{\hyperlink{cite.yang2025pensievediscuss}{\makecell{Pensieve\\Discuss}}} &
\textbf{\hyperlink{cite.taneja2024jillwatson}{\makecell{Jill\\Watson}}} &
\textbf{\hyperlink{cite.dai2025parrot}{Parrot}} &
\textbf{\framework (Ours)} \\
\hline
\makecell[l]{Guides Instead of\\Giving Answers}
      & $\checkmark$ & $\bullet$ & $\bullet$ & $\times$ & $\checkmark$ \\ \hline
\makecell[l]{Handles Coding\\Help}
      & $\bullet$ & $\checkmark$ & $\times$ & $\times$ & $\checkmark$ \\ \hline
\makecell[l]{Handles Theory\\Questions}
      & $\checkmark$ & $\bullet$ & $\checkmark$ & $\checkmark$ & $\checkmark$ \\ \hline
Generates Quizzes
      & $\bullet$ & $\times$ & $\times$ & $\times$ & $\checkmark$ \\ \hline
\makecell[l]{Checks Quiz\\Answers}
      & $\bullet$ & $\times$ & $\times$ & $\times$ & $\checkmark$ \\ \hline
\makecell[l]{Uses Retrieval\\to Stay Grounded}
      & $\bullet$ & $\times$ & $\checkmark$ & $\checkmark$ & $\checkmark$ \\ \hline
\makecell[l]{Late \& Semantic\\Chunking}
      & $\times$ & $\times$ & $\times$ & $\times$ & $\checkmark$ \\ \hline
Agentic Pipeline
      & $\bullet$ & $\times$ & $\bullet$ & $\checkmark$ & $\checkmark$ \\ \hline
\makecell[l]{Adaptivity to\\Student Attempts}
      & $\checkmark$ & $\bullet$ & $\checkmark$ & $\checkmark$ & $\checkmark$ \\ \hline
\makecell[l]{One System\\Covers All Tasks}
      & $\times$ & $\times$ & $\times$ & $\times$ & $\checkmark$ \\
\hline
\end{tabularx}
\end{table}

Table \ref{tab:ablation} contextualizes \framework against representative LLM-based tutoring systems across pedagogical and technical dimensions. In guidance style, Khanmigo emphasizes Socratic questioning to elicit understanding, Pensieve Discuss provides collaborative support through synchronized editing, and Jill Watson and Parrot deliver direct responses, while \framework employs stepwise scaffolding that decomposes tasks into manageable increments with iterative feedback. For coding support, only Pensieve Discuss and \framework offer dedicated programming environments; however, \framework uniquely implements a generate-execute-verify loop operating at individual step granularity with sandboxed validation, whereas Pensieve Discuss focuses on collaborative editing with autograding at submission level. Most systems handle conceptual Q\&A, but \framework distinguishes itself by routing queries through task-specific retrieval combining late chunking for semantic-lexical fidelity and BM25-FAISS hybrid search with cross-encoder reranking, contrasting with the generic retrieval in Jill Watson and Parrot or the absence of retrieval-grounded responses in Khanmigo and Pensieve Discuss. For formative assessment, only Khanmigo and \framework support quiz generation, though \framework extends this with Bloom-level tagging, distractor rationales, multi-concept coverage validation, and adaptive difficulty progression based on learner performance. Regarding technical architecture, \framework maintains dual vector databases optimized for distinct retrieval objectives (grounding versus diversity), employs late and semantic chunking strategies absent in other systems, and integrates routing, retrieval, quiz synthesis, and coding guidance within a unified agentic pipeline, enabling seamless transitions between conceptual learning, assessment, and applied problem-solving without external tool switching or manual orchestration.

\end{document}